\documentclass[aps,article,twocolumn,superscriptaddress]{revtex4}
\usepackage{graphicx}
\usepackage{epsfig} 
\usepackage{wrapfig}

\usepackage{amsmath}
\usepackage{amsfonts}
\usepackage{gensymb}

\usepackage{url}

\newcommand{\bfB}{{\mathbf{B}}}

\newcommand{\bfE}{{\mathbf{E}}}
\newcommand{\bfe}{{\mathbf{e}}}

\newcommand{\bfk}{{\mathbf{k}}}

\newcommand{\bfr}{{\mathbf{r}}}

\newcommand{\bfv}{{\mathbf{v}}}

\newcommand{\rmb}{{\mathrm b}}
\newcommand{\rmc}{{\mathrm c}}

\newcommand{\rmd}{{\mathrm d}}

\newcommand{\rme}{{\mathrm e}}

\newcommand{\rmi}{{\mathrm i}}

\newcommand{\rms}{{\mathrm s}}

\newcommand{\rmDe}{{\mathrm{De}}}

\newcommand{\rmce}{{\mathrm{ce}}}

\newcommand{\rmpe}{{\mathrm{pe}}}
\newcommand{\rmpi}{{\mathrm{pi}}}

\begin{document}
\title{Cross-field chaotic transport of electrons \\ 
        by E $\times$ B electron drift instability in Hall thruster}

\author{{D. Mandal}}
\affiliation{Aix-Marseille Universit{\'e}, CNRS, UMR 7345-PIIM Laboratory, Marseille, France}
\affiliation{ Indo-French Centre for the Promotion of Advanced Research-CEFIPRA, New Delhi, India}
\author{Y. Elskens}
\affiliation{Aix-Marseille Universit{\'e}, CNRS, UMR 7345-PIIM Laboratory, Marseille, France}
\author{N. Lemoine}
\affiliation{Universit\'{e} de Lorraine, Institut Jean Lamour, UMR-7198, CNRS, France}
\author{F. Doveil}
\affiliation{Aix-Marseille Universit{\'e}, CNRS, UMR 7345-PIIM Laboratory, Marseille, France}

\begin{abstract}
A model calculation is presented to characterize the anomalous cross-field 
transport of electrons in a Hall thruster geometry. The anomalous nature of 
the transport is attributed to the chaotic dynamics of the electrons arising 
from their interaction with fluctuating unstable electrostatic fields of the 
electron cyclotron drift instability that is endemic in these devices. 
Electrons gain energy from these background waves leading to a significant 
increase in their temperature along the perpendicular direction 
$T_\perp / T_\parallel \sim 4$ and an enhanced cross-field electron transport 
along the thruster axial direction. It is shown that the wave-particle 
interaction induces a mean velocity of the electrons along the axial direction, 
which is of the same order of magnitude as seen in experimental observations.

\bigskip


   52.25.Fi   Transport properties \\
   52.75.Di   Ion and plasma propulsion  \\

\end{abstract}


\maketitle

\section{Introduction}
\label{sec:intro}
%
Hall thrusters \cite{Morozov:a} are gridless ion sources that are frequently 
used as space
propulsion devices in geostationary satellites and long range missions such
as Earth to Moon missions. They have been the subject of many past 
studies \cite{Lafleur:t,  Adam:j, BoeufGarrigues, Marusov:n, Tsikata, Smirnov:a, Janes}.
A salient feature observed in such studies is the presence of a 
strong cross-field anomalous electron transport along the axial direction of 
the thruster. This has been consistently observed both in model numerical 
simulations
\cite{Lafleur:t, Adam:j, BoeufGarrigues} as well as in laboratory experiments
\cite{Janes, Tsikata, Smirnov:a}, 
and a detailed understanding of this anomalous transport process is still lacking.  
Since the efficiency of the
thruster decreases with an increase in the anomalous electron transport
\cite{Smirnov:a}, it is important to gain some understanding of the underlying 
mechanism driving such a transport. 

Our present work is motivated by a desire to throw some light on this process, 
and we attempt to do so by analyzing the characteristics of this transport and 
developing a physics model to describe the origin of the transport. 
Since the ionization efficiency in the thruster chamber is more than 
90$\%$, the density of neutral atoms is so low that electron collisions 
cannot explain the high electron flux observed experimentally. 
Indeed, the electron transport coefficients are 100 times larger
than those given by the collisional transport model \cite{AdamBoeuf:jcjp}.
Since the collisional transport fails to explain 
the observed cross-field electron transport after the channel exit, 
other explanations have been proposed in the past. Among them the 
non-collisional transport due to the interaction of electrons with the 
electric fields of the numerous electrostatic instabilities that can occur 
is an attractive candidate. Indeed, 2D (azimuthal and axial) PIC 
simulations \cite{Adam:j} show that turbulence alone (without 
any wall conductivity that could not be modeled in this simulation) is able 
to drive a high enough electron transport to explain anomalous transport. The 
dominant instability seen in those simulations was also observed experimentally 
\cite{Tsikata} and identified theoretically as the $\bfE \times \bfB$ electron 
drift instability \cite{Cavalier}.

The $\bfE \times \bfB$ electron drift instability, also called the electron 
cyclotron drift instability or beam cyclotron instability \cite{Gary:s}, is 
observed in a magnetized plasma under conditions when the ion motion is hardly 
modified by the magnetic field whereas the electrons experience a strong 
drift, resulting in a huge velocity difference between electrons and ions. 
The frequency of this instability is much lower than the electron cyclotron 
frequency ($\omega  \ll  \omega_\rmc$). Therefore, the resonance condition 
with the cyclotron harmonics, 
$\omega -k_\parallel v_\parallel = n \omega_\rmc$ is 
not satisfied. The frequency is of the order of the ion acoustic wave 
frequency. 

The mechanism of the instability is the following. Bernstein waves (whose 
frequencies are multiples of the electron cyclotron frequency) are 
Doppler-shifted towards low frequencies by the high electron drift velocity 
and reach the ion acoustic wave range. The instability occurs when the two 
modes merge \cite{GarySanderson:pj}. The magnetic field and the electron drift 
velocity are the main sources of the $\bfE \times \bfB$ electron drift 
instability. Plasma density, temperature and magnetic field gradients as well 
as ion flows can also play a role \cite{Mikhailovskii}. This instability is 
observed in many magnetized plasma devices like magnetrons for material 
processing \cite{Abolmasov:s}, magnetic filters \cite{BoeufClauster}, 
Penning gauges \cite{Ellison:c}, linear magnetized 
plasma devices dedicated to study cross-field plasma instabilities 
\cite{Matsukuma:m}, Hall thrusters \cite{Morozov:a} and many fusion devices.

The transport resulting directly from this instability has not been quantified 
yet and the mechanism of the instability-electron interaction in this case has 
not been studied. This paper proposes a first investigation into those 
questions based on a simple model calculation.
In particular, we study the electrons dynamics in a slowly time varying 
($\omega \ll \omega_\rmc$)
potential profile in the presence of a constant axial electric field and 
a radial magnetic field. The ion dynamics and their effect 
on electrons are not considered in this model, and in that sense in our model 
the system is not self-consistent.

The paper is organized as follows. In section \ref{sec:model}, we briefly 
describe the Hall thruster mechanism 
and the model considered for the wave dispersion relation and spectrum. 
In section \ref{sec:numerical}, the numerical scheme used for particle trajectory integration 
is detailed. In section \ref{sec:1wave}, we study the behavior of an electron 
interacting with only one Fourier mode fulfilling the instability dispersion 
relation of the $\bfE \times \bfB$ electron drift instability. In section 
\ref{sec:3waves}, we study the behavior of an electron interacting with three 
Fourier modes. In section \ref{sec:conclusion}, we show that due to 
the strong wave-particle interactions, the dynamics of each electron becomes 
chaotic, and in the presence of more than one wave, we find a significant 
amount of cross-field electron transport along the axial direction. 
%
\section{Elementary model}
\label{sec:model}
\begin{figure}
\includegraphics[width=8 cm] {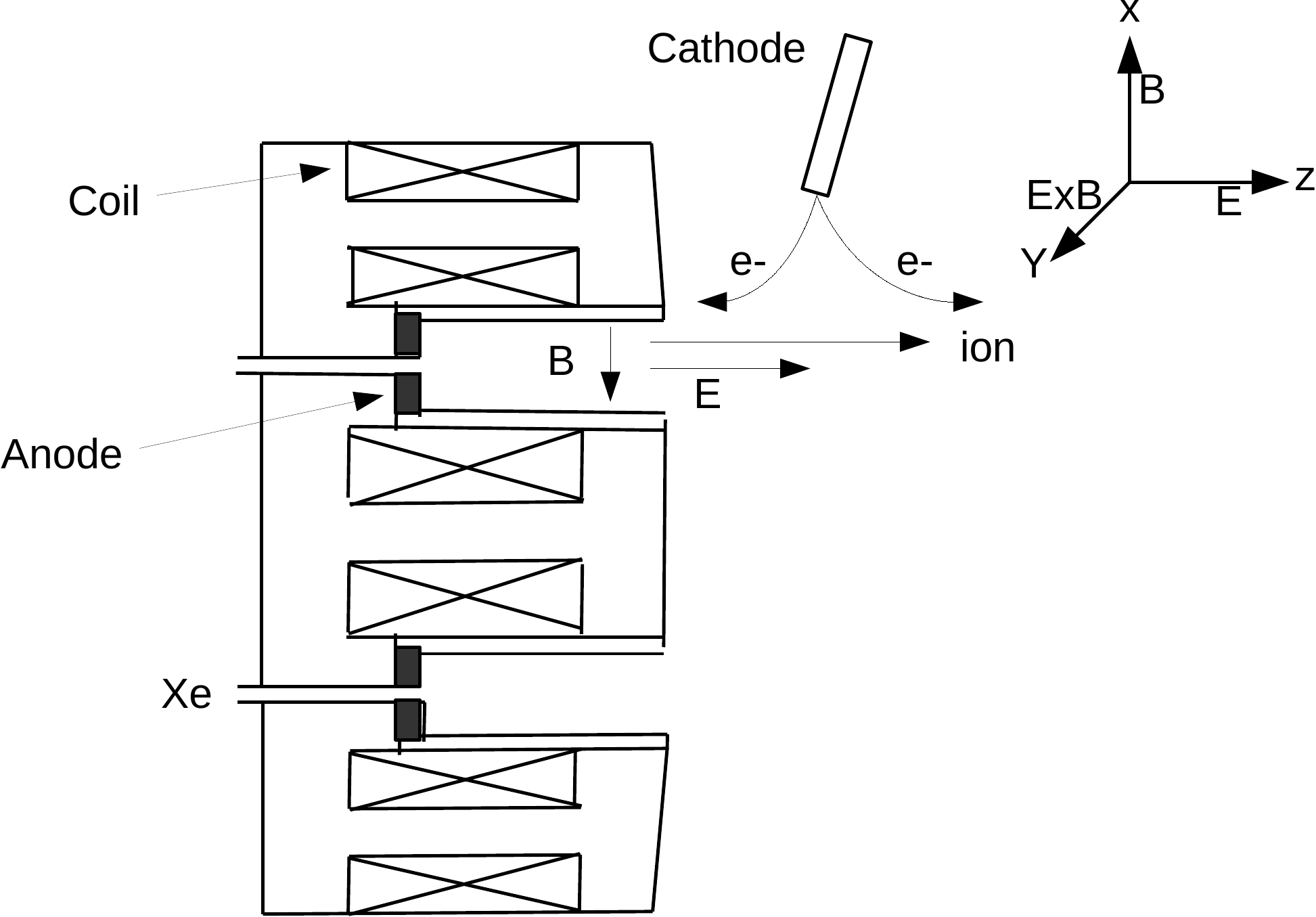}
\caption{Schematic diagram of Hall thruster. 
$B$ is the magnetic field, $E_0$ is the constant axial electric field, $e^-$ 
denotes electrons, and the top-right sketch presents the Cartesian 
coordinates for our numerical simulation.}
\label{Thruster_sch}
\end{figure}

In a Hall thruster, plasma is formed between two co-axial dielectric cylinders. 
Electrons are injected from an emissive cathode placed outside the 
exhaust plane and, due to the presence of the strong radial magnetic field, 
these electrons start to gyrate around magnetic lines and become magnetized.
The combination of the axial electric field and the radial magnetic field 
generates a strong $\bfE \times \bfB$ drift motion in the azimuthal direction.
This creates closed Hall current loops. The magnetized electrons are trapped 
in this configuration and stay for a long time within the channel. This 
results in a decrease of the electron conductivity in the axial direction.
Xenon atoms injected through the anode at the end of the channel are ionized 
by the electrons drifting at a high velocity. Since the ions are not 
magnetized, they are extracted from the plasma and the axial electric field 
accelerates them from the ionization region without collision, as sketched  
in Fig.~\ref{Thruster_sch}. The electrons injected from an emissive 
cathode help to generate the plasma and also help to neutralize the ion beam.
We consider a Cartesian coordinate system for the numerical modeling, 
with the $x$-direction as the magnetic field direction, the $y$-direction as 
the $\bfE \times \bfB$ drift direction and the $z$-direction as the constant 
electric field direction, representing the radial, 
azimuthal and axial directions respectively of the thruster chamber. 
Fig.~\ref{Thruster_sch} presents these three directions. 

In the context of a Hall thruster, using a cold fluid equation for unmagnetized 
ions and a Vlasov kinetic equation for magnetized electrons, 
Cavalier \textit{et al.} \cite{Cavalier} derived 
a 3D dispersion relation for the $\bfE \times \bfB$ drift instability in the form
\begin{eqnarray}\nonumber
  1 + k^2 \lambda_\rmDe^2
     + g\left(\frac{\omega - k_{y} v_\rmd}{\omega_\rmce}, 
                 (k_{x}^2 + k_{z}^2) \rho_\rme^2,
                 k_{x}^2 \rho_\rme^2 \right) 
  &&  \\
  - \frac{k^2 \lambda^{2}_\rmDe \omega^{2}_\rmpi} {(\omega - k_{z} v_{\mathrm{i,b}})^2} 
& = & 0, \qquad
\label{dispersion_rel}
\end{eqnarray}
where $\lambda_\rmDe$ is the electron Debye length, 
$v_\rmd = E_z / B$ is the electron  drift velocity, 
$v_{\mathrm{i,b}}$ is the ion beam velocity, 
$\rho_\rme=v_{\mathrm{the}}/\Omega_\rmce$ is the electron Larmor radius,
$v_{\mathrm{the}}$ is the electron thermal velocity ; 
$\omega$, $\omega_\rmce$ and $\omega_\rmpi$ 
are the mode, electron cyclotron and ion plasma frequency, respectively, 
while $k_{x}$, $k_{y}$, $k_{z}$ and $k$ are the $x$, $y$ and $z$ 
components and modulus of wave vector $\bfk$, respectively. 
$g$ is the Gordeev function \cite{Gordeev}: 
$g(\Omega,X,Y) = \frac{\omega}{2Y} \exp(-X) 
\sum_{m=0}^{\infty}Z(\frac{\Omega-m}{\sqrt{2Y}})I_{m}(X)$
where $Z(x)$ is the plasma dispersion function 
and $I_{m}$ is the modified Bessel function of first kind. 
This instability described by Eq.~(\ref{dispersion_rel}) can 
grow to a sufficient level of turbulence into a non-magnetic ion-acoustic
instability with modified angular frequency and growth rate \cite{BoeufGarrigues}
\begin{eqnarray}
  \omega 
  & \approx & 
  k_{x}v_{\rm {i,b}} + \frac{k c_\rms}{\sqrt{1 + k^2\lambda_\rmDe^2}}, 
  \nonumber
  \\
  \gamma 
  & \approx & 
  \sqrt{\frac{\pi m_\rme}{8m_\rmi}} \frac{k_{y} v_\rmd} {(1 + k^2\lambda_\rmDe^2)^{3/2}}, 
\label{modfied_DR}
\end{eqnarray}
respectively, where $c_\rms$ is the ion acoustic velocity.
\begin{figure}
\includegraphics[width=8 cm]{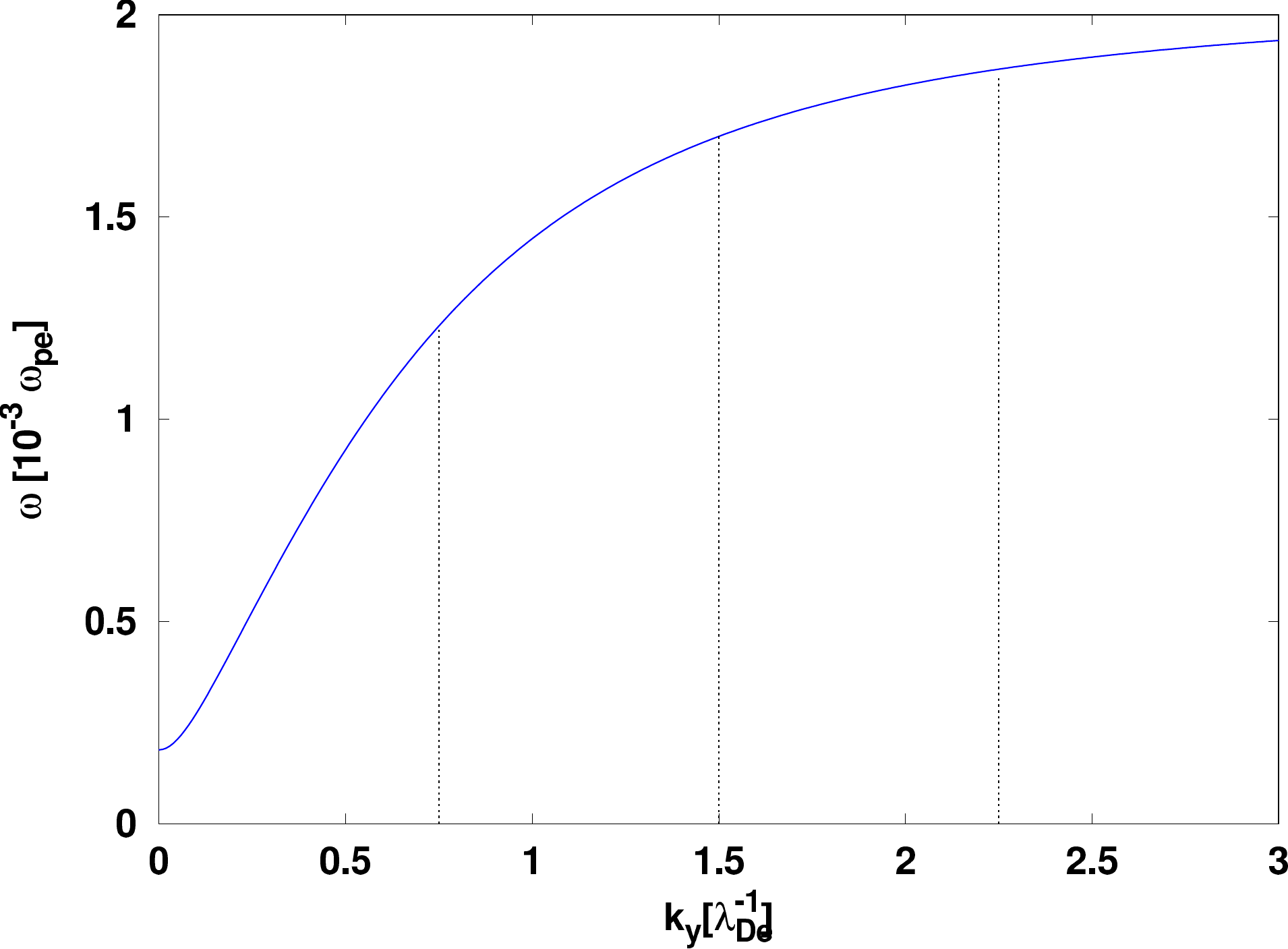}
\caption{Solutions of the 3D $\bfE \times \bfB$ electron drift instability 
for $k_x = 0.03$. The three vertical lines locate the three modes which 
are taken for our numerical study.}
\label{dispersion_plot}
\end{figure}
This analytical model for the dispersion relation fits well with experimental 
data. We consider a constant electric field $\bfE_0 = E_0 \, \hat \bfe_z$
along the $z$-direction and a constant magnetic field 
$\bfB = B_0 \, \hat \bfe_x$ along $x$-direction.

Experimentally, the observed propagation angle of the instability-generated 
wave deviates by $\tan^{-1} ({ k_z / k_y }) \sim 10-15\,\degree$  from the
azimuthal $y$-direction near the thruster exit plane. Further from the exit 
plane, the propagation becomes progressively more azimuthal \cite{Tsikata}. 
Hence, the wave vector along the axial direction $k_z \sim 0.2 \, k_y$, and the 
electric field along the axial direction is dominated by the stronger constant 
field $E_0 \, \hat \bfe_z$. Therefore for simplicity, we consider that the 
unstable modes are confined in $x-y$ (i.e., $r - \theta$) plane only. 
Then, the time varying part of the potential in $x-y$ plane 
is constructed as a sum of unstable modes. 
The total electric field acting on the particle is
\begin{eqnarray} 
  \bfE(x,y,z,t) 
  & = & \sum_n (k_{n x} \hat \bfe_x + k_{n y} \hat \bfe_y) \, {\phi}_{0 n} \sin \alpha_n(x,y,t) 
  \nonumber \\ 
  && +  E_0 \, \hat \bfe_z,
\label{e-field}
\end{eqnarray} 
with local phase 
$\alpha_n (x,y,t) = k_{n x} x +  k_{n y} y - \omega_n t + \zeta_n$, 
where $n$ is a label for different modes with wave vector $\bfk_n$, 
angular frequency $\omega_n$ and phase ${\zeta_n}$. $\bfk_n$, 
$\omega_n$ follow the dispersion relation eq.~(\ref{dispersion_rel}) and 
phases  $\zeta_n$ are random. 
Here, the position $\bfr$, velocity $\bfv$, time $t$, and potential $\phi_0$ 
are normalized with Debye length $\lambda_\rmDe$, 
thermal velocity $v_{\mathrm{the}}$, 
reciprocal $\omega_\rmpe^{-1}$ of the electron plasma frequency, 
and $m_\rme v_{\mathrm{the}}^2 / \vert q_\rme \vert$, 
respectively.
We choose the amplitude ${\phi}_{0n}$ of all the modes equal to 
the saturation potential \cite{BoeufGarrigues} at the exit plane of the thruster
$\vert \delta \phi_{y,{\mathrm{rms}}} \vert =T_\rme / (6\sqrt{2}) = 0.056 \, m_\rme v_{\mathrm{the}}^2$. 

We consider three modes $(n = 1, 2, 3)$ 
with $(k_{nx}, k_{ny}, \omega_n) = (0.03, 0.75, 1.23 \times 10^{-3})$, 
$(0.03, 1.5, 1.7 \times 10^{-3})$ and $(0.03, 2.25, 1.87 \times 10^{-3})$, respectively. 
The location of these three modes is shown in Fig.~\ref{dispersion_plot} by three vertical lines. 
In normalized units, 
$\vert q_\rme \vert B_0 / m_\rme = 0.1 \, \omega_\rmpe$, 
$\vert q_\rme \vert E_0 / m_\rme = 0.04 \, \omega_\rmpe v_{\mathrm{the}}$, 
and ${v_\rmd = 0.4 \, v_{\mathrm{the}}}$.
Therefore, for all three modes, the $y$-component of phase velocity 
$\omega_n / k_{ny} \ll v_\rmd$.

%
\section{Numerical method }
\label{sec:numerical}
The equations of motion of the particle are
\begin{eqnarray}
  \frac{\rmd \bfr}{\rmd t} = \bfv,
  ~~~
  \frac{\rmd \bfv}{\rmd t} = \frac{q_\rme}{m_\rme} (\bfE + \bfv \times \bfB).
  \label{eq_motion}
\end{eqnarray} 
Because $\bfE$ depends on space, the infinitesimal generators for both 
equations do not commute, and one uses a time-splitting numerical integration scheme. 
The first equation is integrated in the form 
$\bfr (t + \Delta t) = {\cal T}_{v, \Delta t} (\bfr(t)) =  \bfr (t) + 
\bfv \Delta t$.
For the second equation, we separate the electric integration
$\bfv (t + \Delta t) = {\cal T}_{E, \Delta t} (\bfv(t)) 
= \bfv (t) + (q_\rme / m_\rme) \bfE \Delta t$
from the magnetic integration, which solves only the gyro-motion. 
For the latter, we use the Boris method \cite{boris}, formally
$\bfv(t + \Delta t) = {\cal T}_{B,\Delta t} \bfv (t)$.
As a result, we use a second-order symmetric scheme
\begin{equation}
  \left( \begin{array}{c} \bfr (t+\Delta t) \\ \bfv (t+\Delta t) \end{array} \right) 
  = \mathcal{A}
     \left( \begin{array}{c} \bfr (t) \\ \bfv (t) \end{array} \right),
  \label{matrix_oper}
\end{equation}
with the nonlinear map
\begin{equation}
\mathcal{A} 
= {\cal T}_{v, \Delta t/2} \circ {\cal T}_{E, \Delta t/2} 
                                      \circ {\cal T}_{B, \Delta t} 
                                      \circ {\cal T}_{E, \Delta t/2} 
                                      \circ {\cal T}_{v, \Delta t/2}  .
\label{Sympl_opert}
\end{equation}
\begin{figure}
\includegraphics[width=0.85\linewidth]{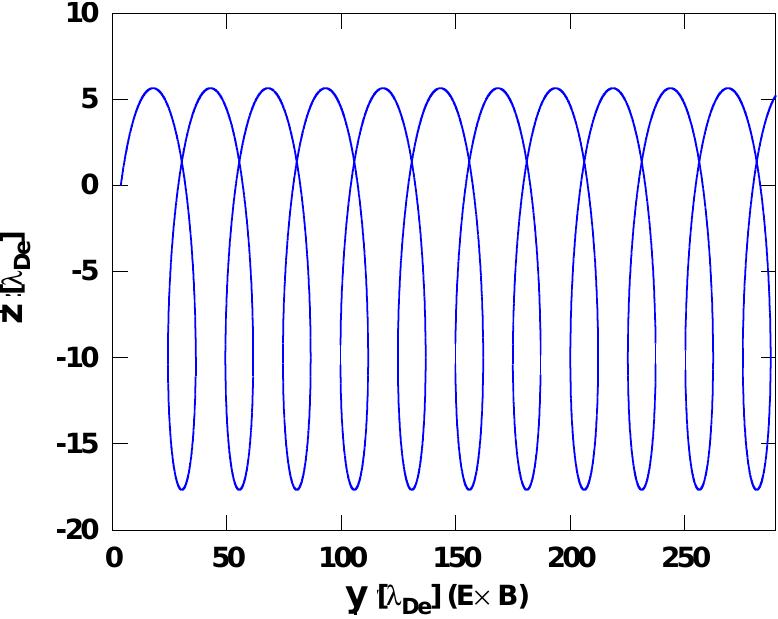}
\caption{
Regular trajectory of a single particle in presence of a 
constant magnetic field $\omega_\rmc = 0.1 \, \omega_\rmpe$ 
and a constant electric  field $E_0 = 0.04$. 
}
\label{regular_traj}
\end{figure}
\begin{figure}
\includegraphics[width=0.85 \linewidth]{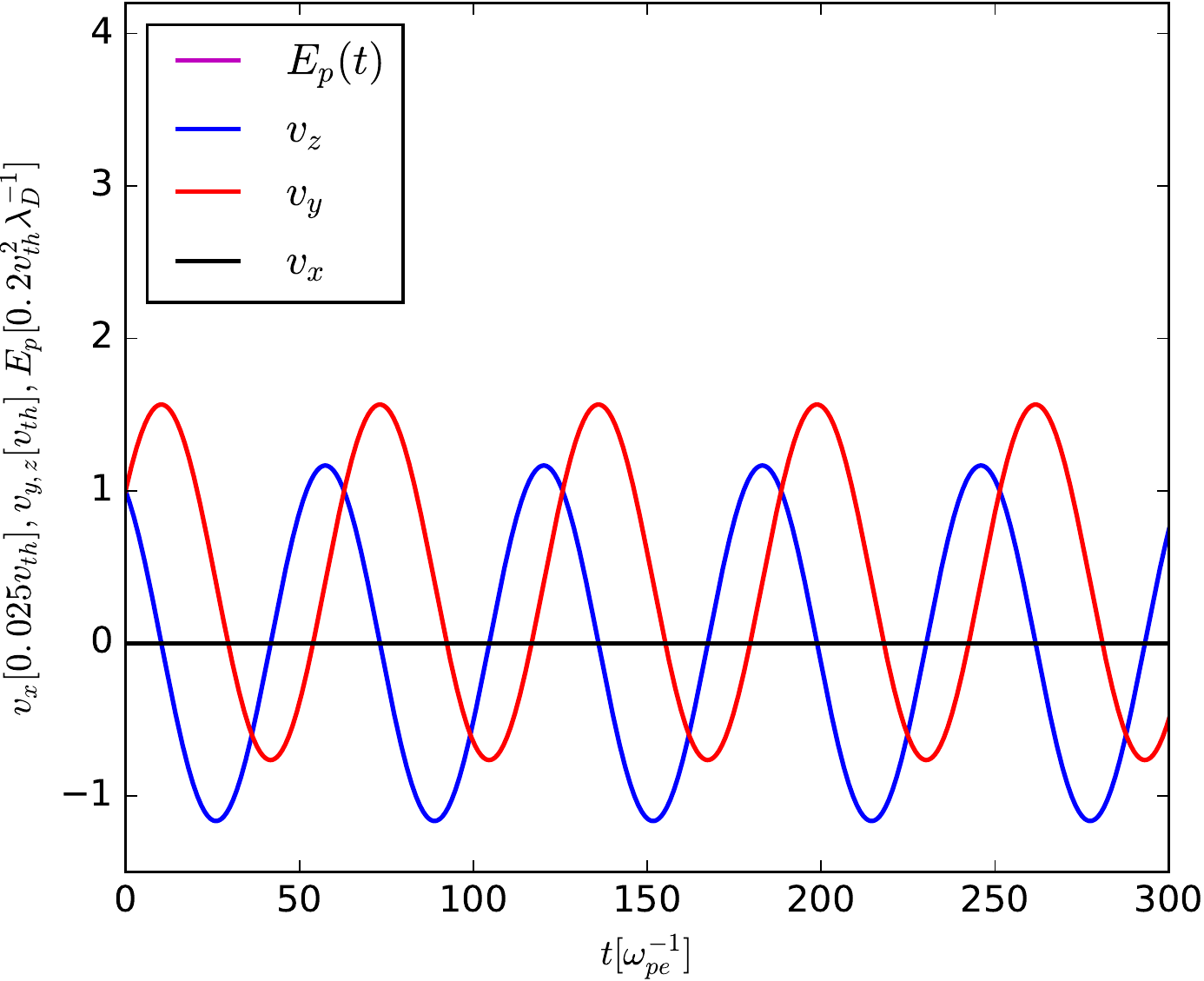}
\caption{
Three components of velocity $v_x$ (black), $v_y$ (red), 
$v_z$ (blue) and the $y$-component of electric field at the particle location 
$E_{\mathrm{p}}(t)$ (magenta).
Since there is no background electrostatic wave, 
the electric field amplitude at the particle location vanishes and the 
magenta line coincides with the black one. 
}
\label{regular_velocity}
\end{figure}
To understand the effect of waves, 
we first solve the equations of motion Eq.~(\ref{eq_motion}) numerically 
for a single particle trajectory with initial velocity 
$v_{0x} = 0, v_{0y} = 1$ and $v_{0z} = 1$ 
in presence of a constant electric field $E_0 = 0.4$ along $z$-direction 
and a constant magnetic field along $x$-direction 
such that $\omega_\rmc = 0.1 \, \omega_\rmpe$. 
Since there is no background electrostatic wave ($E_x = E_y = 0$), 
the particle exhibits regular cycloid motion. 
Therefore, the position co-ordinates $x, y, z$ follow the relation 
$(y - c - a \tau)^2 + (z - b)^2 = c^2 + b^2$, where $a = v_\rmd / \omega_\rmc$,
$b = (v_\rmd - v_{0y}) / \omega_\rmc$, $c = v_{0z} / \omega_\rmc$ and 
$\tau = \omega_\rmc t$ and the velocity components are 
$v_x = v_{0x}, v_{y}= {\rm v_{\perp 0}} \cos(\tau) + v_\rmd$ and 
$v_{z} = {\rm v_{\perp 0}} \sin(\tau)$, where 
${\rm v_{\perp 0}} = \sqrt{v_{0z}^2 + (v_{0y} - v_\rmd)^2}$ and 
$(v_{0x}, v_{0y}, v_{0z})$ are the initial velocity components. 
Figs~\ref{regular_traj} and \ref{regular_velocity} present the trajectory and the velocity components 
of the particle. Along the $y$-direction, there is a drift velocity 
$v_\rmd = 0.4 \, v_{\mathrm{the}}$. 
Since $v_x = 0$, the trajectory is confined in the $y-z$ plane.    
%
\section{Particle trajectory in presence of one wave}
\label{sec:1wave}
%
\begin{figure}[h!]
\includegraphics[width= \linewidth]{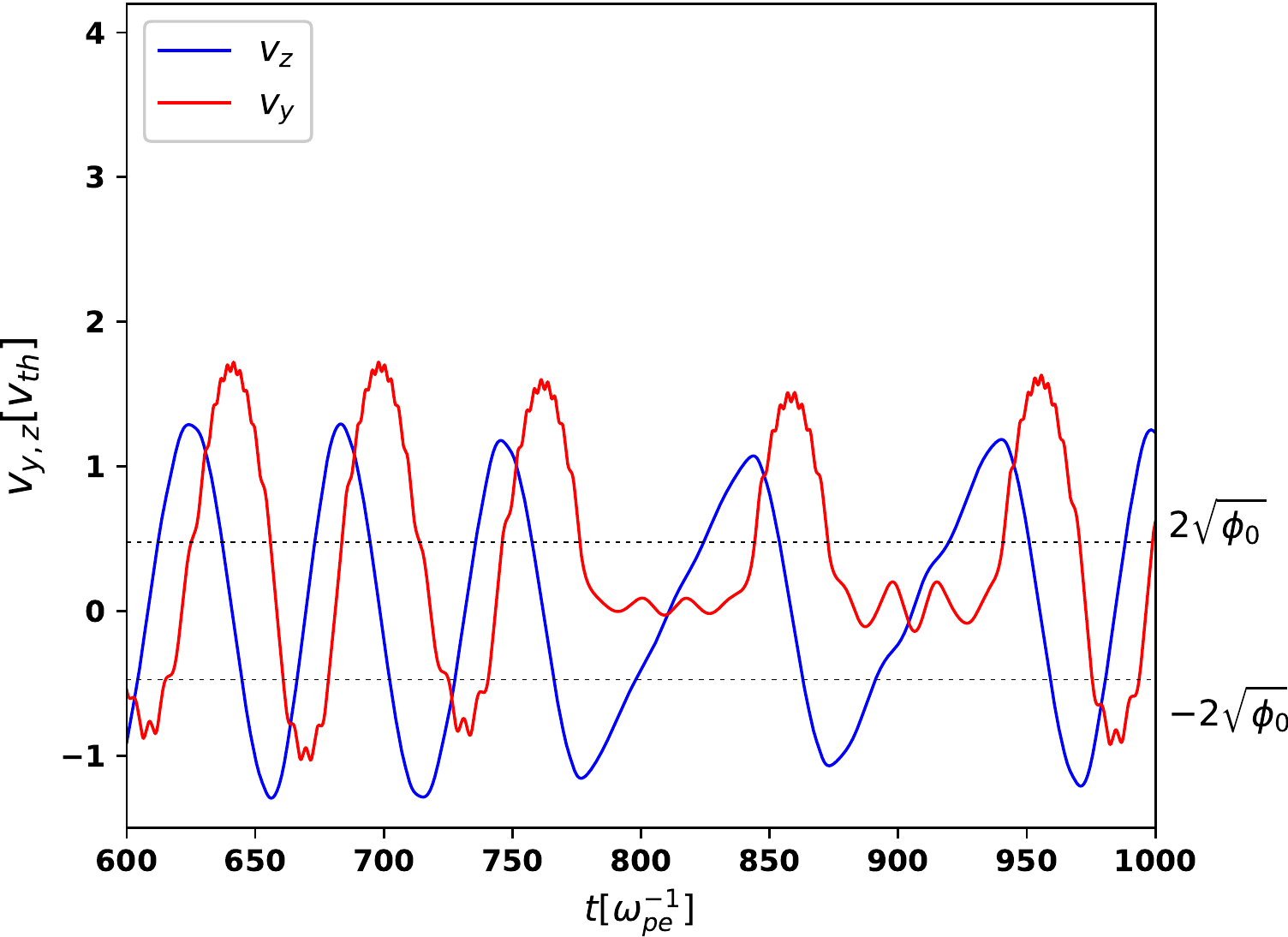}
\caption{Particle evolution in the presence of a single background 
electrostatic wave with $n = 2$. Velocity components $v_y$ (red) 
and $v_z$ (blue) of one particle. Near $t = 800$ and $900$, the particle is 
trapped in the wave potential and it oscillates with the time period
$\tau_\rmb = 18 \, \omega_\rmpe^{-1}$. }
\label{vy_vz_chaos}
\end{figure}
\begin{figure}[h!]
\includegraphics[width= \linewidth]{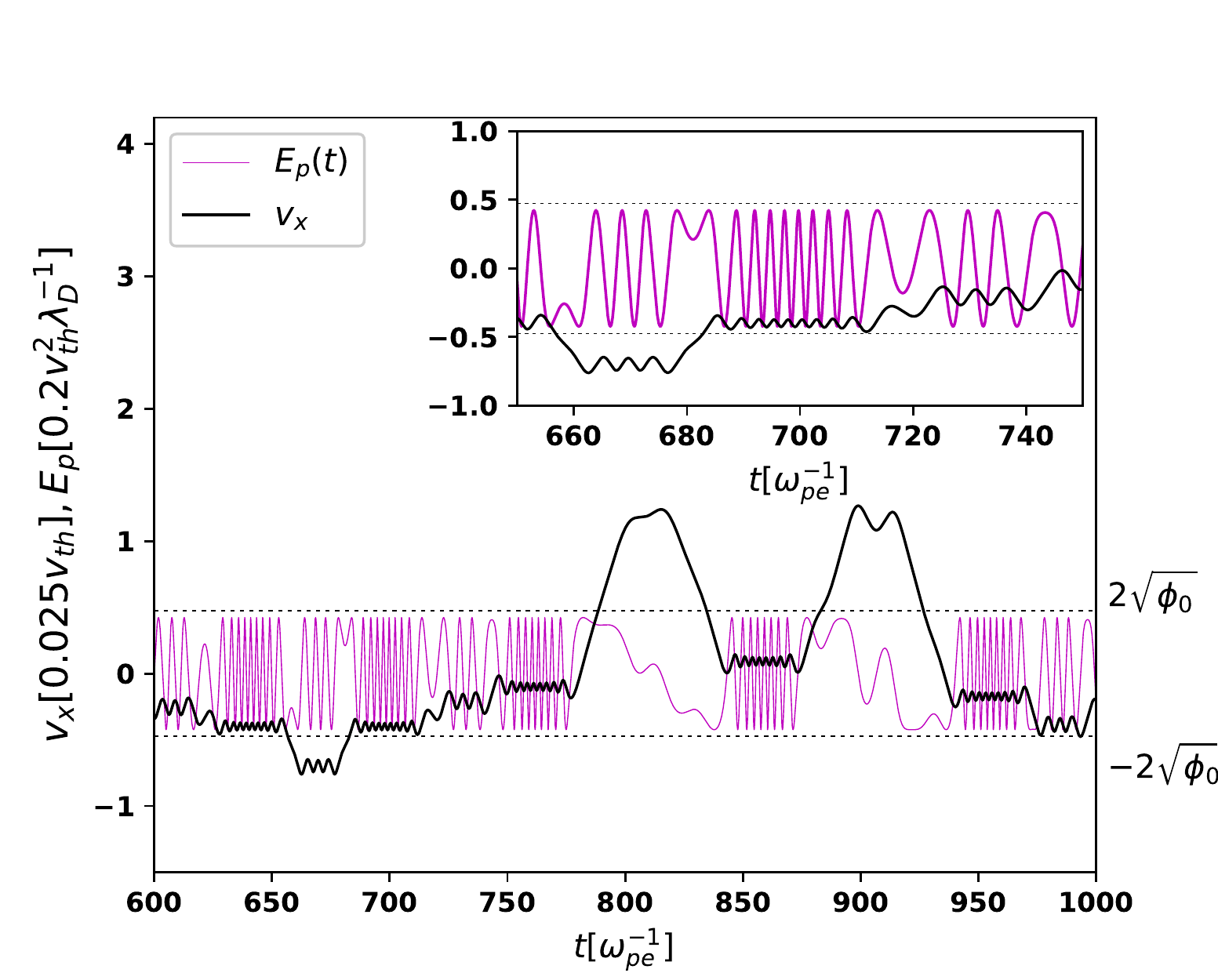}
\caption{
Particle evolution in the presence of a single background 
electrostatic wave with $n = 2$. 
$v_x$ (black solid line), electric field at particle location 
$E_{\mathrm{p}}(t)$ (magenta line). Black dotted horizontal lines show the 
location of $\pm 2 \sqrt{\phi_0}$.
}
\label{vx_Ex_chaos}
\end{figure}
In the presence of a background electrostatic wave, 
the wave-particle interaction modifies the cyclotron motion. 
The strength of the wave-particle interaction depends on the 
wave amplitude and the particle velocity. 
Fig.~\ref{vy_vz_chaos} presents the time evolution of $v_y$ (red line) and 
$v_z$ (blue line), and Fig.~\ref{vx_Ex_chaos} presents the time evolution 
of $v_x$ (black line) and the $y$-component $E_{\mathrm{p}}(t)$ of electric 
field at particle location (magenta line). Due to the cyclotron motion, 
$v_y$ oscillates about the drift  velocity $v_\rmd = 0.4$. During each 
cyclotron oscillation, when $| v_y | \leq 2 \sqrt{\phi_0}$ (denoted by 
black dashed lines) the particle interacts strongly with the electrostatic 
wave, and the electric field $E_{\mathrm{p}}(t)$ enhances/reduces the $v_x$ 
value by a large amount. 

The inset of Fig.~\ref{vx_Ex_chaos} presents, during a strong interaction, 
according to the sign of $E_{\mathrm{p}}$, jumps of $v_x$ in positive and 
negative direction. Moreover, during this strong interaction depending on 
the local potential profile, the particle may be trapped in the wave potential 
well and oscillate with the bounce frequency 
${\omega_\rmb = 0.35 \, \omega_\rmpe}$. 
In Fig.~\ref{vy_vz_chaos} near $t=800$ and $900$, it is trapped. 
One essential condition for the trapping is ${\omega_\rmb > \omega_\rmc }$, 
where $ \omega_\rmb = k_y \sqrt{| q_\rme | \phi_0 / m_\rme}$ is the bounce 
frequency. Since $k_y \gg k_x$, the condition for trapping is easily satisfied 
along the $y$-direction, therefore the particle bounces back and forth along 
the $y$-direction and moves freely along the $x$-direction. Hence, along 
the $x$-direction it gains/loses energy from/to the wave, which causes a large 
change in $v_x$. Finally, depending on the local potential value, 
it may escape from the wave and again start to exhibit cyclotron motion. 
Therefore, the duration of trapping depends on $v_x$ and 
${\omega_\rmb/ \omega_\rmc }$. It is observed that, for small 
$v_x \ll \sqrt{\phi_0}$, this trapping is easily observed 
for ${\omega_\rmb/ \omega_\rmc \geq 2}$. 

Outside the strong interaction 
region, due to the large particle velocity, the electric field at particle 
location $E_{\mathrm{p}}$ changes rapidly, which generates the small-amplitude 
fast oscillation in $v_x$. The component $v_y$ is also modulated due to this 
fast change in $E_{\mathrm{p}}(t)$. Since the electric field along 
$z$-direction $E_0 \, \hat \bfe_z$ is constant, the amplitude of the fast 
oscillation in $v_z$ is negligible. The motion along the $z$-direction is 
coupled with the other two directions due to $\bfv \times \bfB$ term of 
Lorentz force, therefore $v_z$ is also modified during the strong interactions. 
In Fig.~\ref{vy_vz_chaos} at $t= 900$, during trapping, the oscillation of 
$v_z$ is observed with frequency $\omega_\rmb$, on top of cyclotron motion. 
\begin{figure}[h!]
\includegraphics[width= \linewidth]{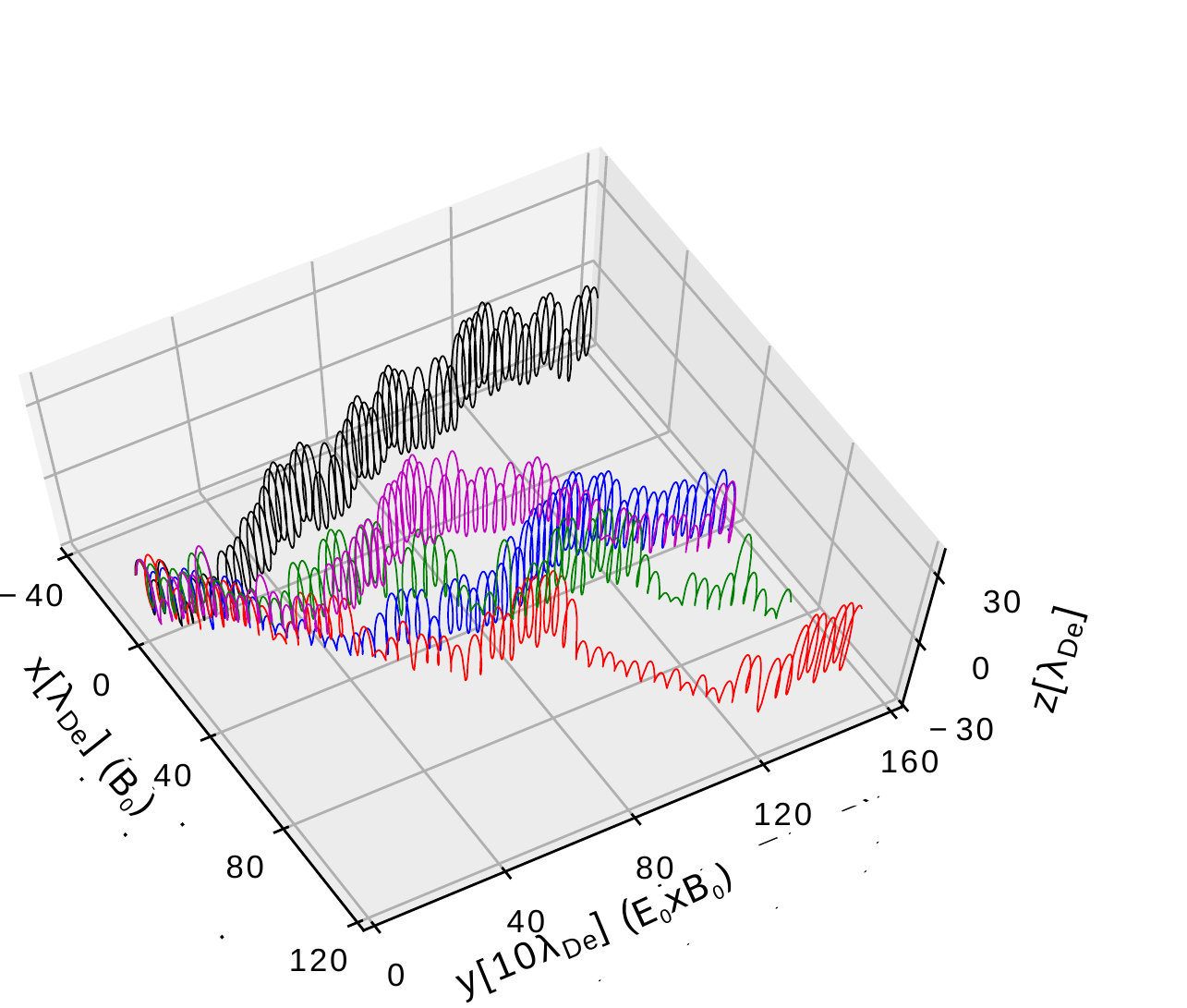}
\caption{Trajectories of 5 different particles with different initial phase 
in the presence of a single background electrostatic wave with $n=2$.}
\label{Traj_chaos}
\end{figure}
\begin{figure}[h!]
\includegraphics[width= \linewidth]{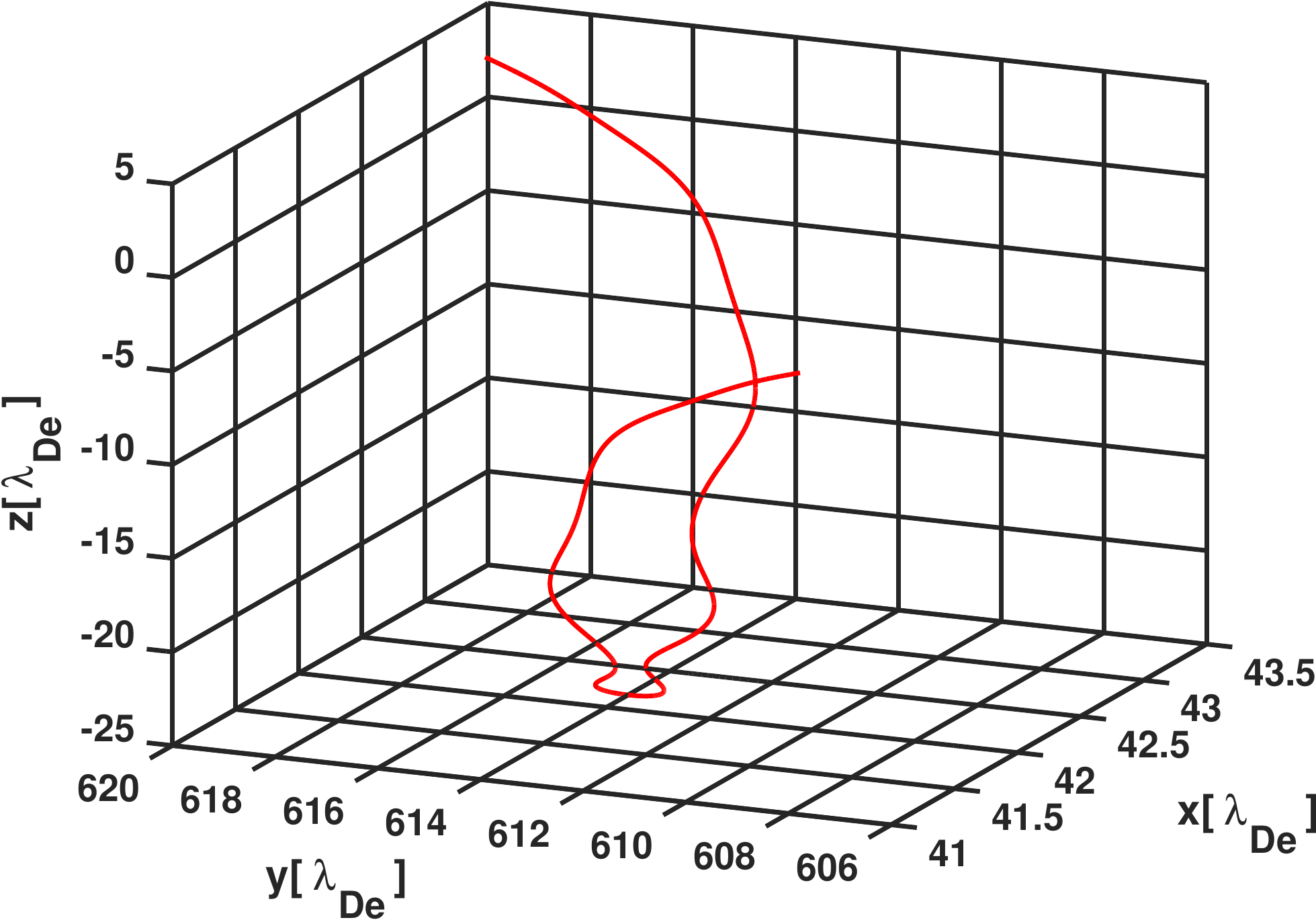}
\caption{
Trajectory of single particle during trapping.
} 
\label{trap_traj}
\end{figure}
\begin{figure}[h!]
\includegraphics[width= \linewidth]{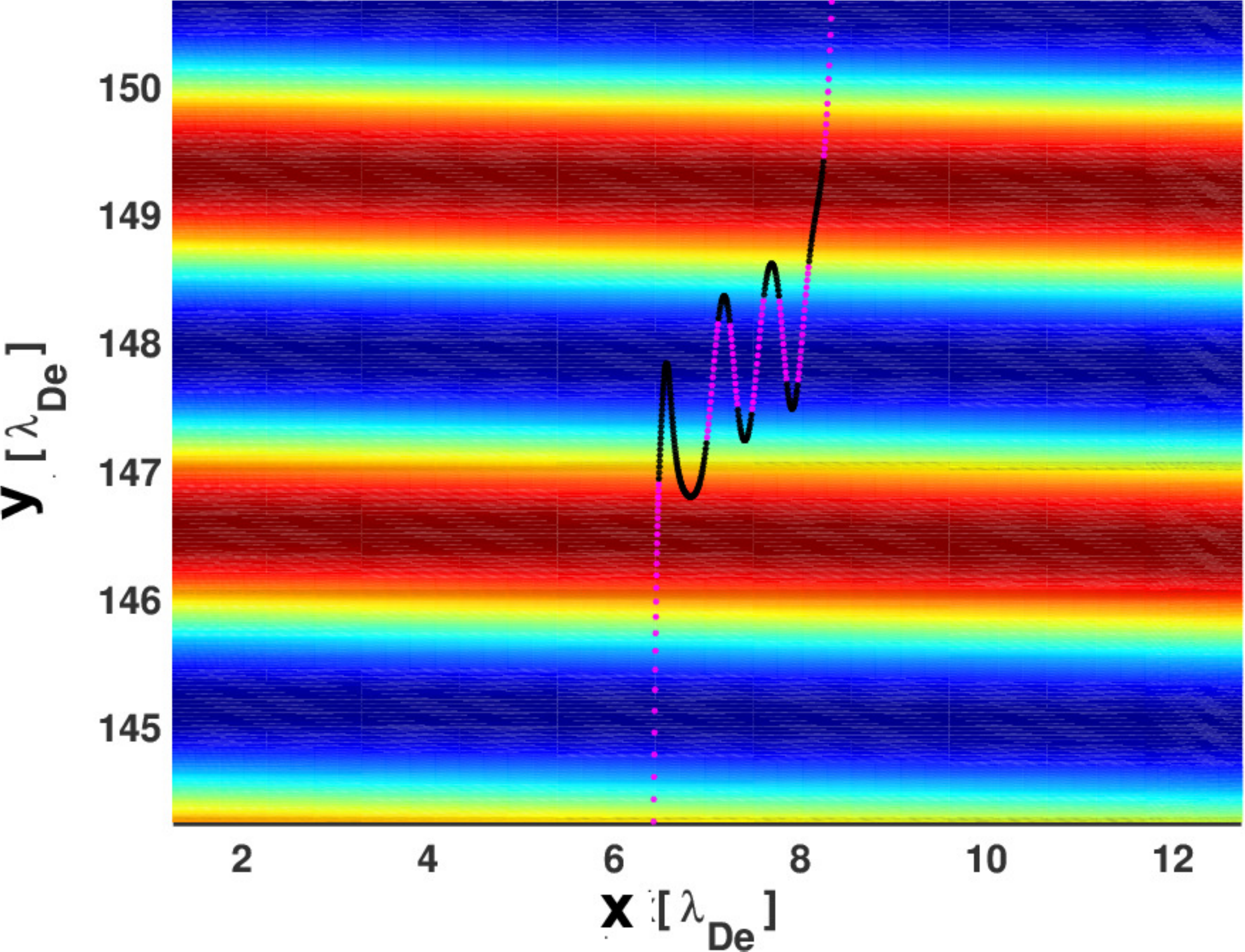}
\caption{
Trapping of particle in the potential well of background wave.
Magenta dots: particle energy higher than the maximum potential energy
$\phi_0$ of the background wave. Black dots: particle energy below $\phi_0$. 
The colour surface plot presents the potential profile of the background wave 
where red denotes the larger values and blue the lower values.
} 
\label{trap_poten}
\end{figure}

Fig.~\ref{Traj_chaos} displays the trajectories of 5 particles with slightly 
different initial positions. 
In the absence of the electrostatic wave, they exhibit cyclotron motion with 
drifting guiding center, and their trajectories remain confined in the $y-z$ 
plane. Due to the strong interaction with the electrostatic wave in 
presence of magnetic field, each trajectory evolves differently 
and they separate exponentially from each other, so that the dynamics becomes 
chaotic. Each strong interaction causes a change in the trajectories along 
$x$, and, depending on the strength of the electric field at particle location, 
$v_y$ may increase or decrease after each strong interaction, which modifies 
the gyroradius ($r_\rmb = v_{\perp} / \omega_\rmc$) accordingly. 

Fig.~\ref{trap_traj} presents a small portion of trajectory during 
trapping. 
Since the particle is trapped along $y$-direction, 
it oscillates within the wavelength 
$\lambda = 2\pi/k_y \sim 4 \, \lambda_\rmDe$ 
and Fig.~\ref{trap_poten} presents the $x-y$ projection of the trajectory 
during trapping. The colour surface plot presents the background wave 
potential. Since $\omega \ll \omega_\rmc$, during the strong interaction
the wave potential remains constant. The magenta dots mark the particle 
location when its energy is greater than the maximum potential energy of the 
electrostatic wave $\phi_0$, and black dots are associated with the particle 
energy below $\phi_0$. During climbing up the potential hill, it loses energy 
and oppositely it gains energy during descent; finally, if, at the top of the 
potential hill (dark red), the particle energy is greater than the potential 
energy $\phi_0$, it detraps from the potential well. Therefore, the trapping 
phenomena depend on the wave potential at the particle location : sometimes 
it may get trapped in the potential well and sometimes it just takes energy 
from the wave and escapes from the potential well. 
During trapping, its average $y$ location remains unchanged. 
Due to this strong wave-particle interaction, the dynamics of the particle 
becomes chaotic. The duration of strong interaction depends on 
$\omega_\rmb / \omega_\rmc$, therefore, for a single wave, chaos will occur 
for amplitudes $\phi_0$ satisfying the inequality 
$\phi_0 \ge \omega_\rmc^2 / k_y^2$. For thruster parameter values, all three 
waves individually satisfy this criterion. 

In the presence of two and three waves, the dynamics becomes more chaotic and 
this threshold value is reduced. With increase of the potential $\phi_0$ and 
the wave vector $k_y$, the bounce frequency of the particle increases, which 
makes the dynamics more chaotic and particles are trapped more frequently in 
the electrostatic wave.
%
\section{Interactions with three waves: Energy gain and axial transport}
\label{sec:3waves}
%
\begin{figure}[b]
\includegraphics[width=\linewidth]{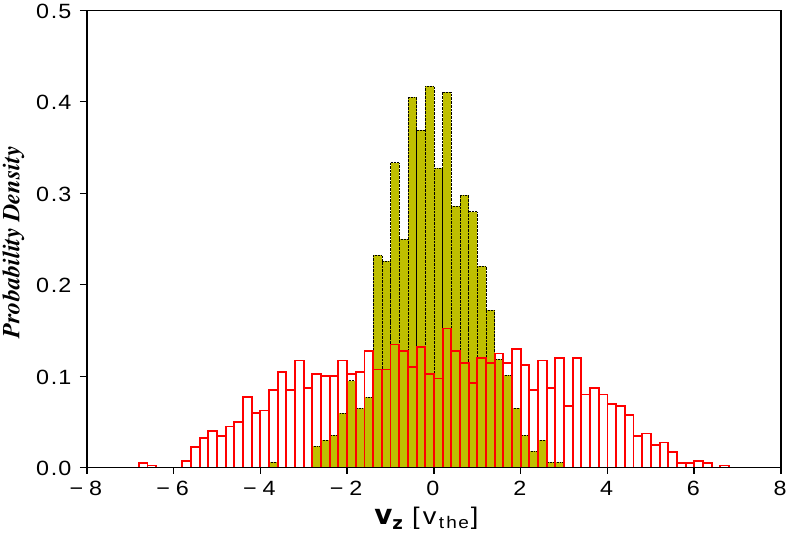}
\caption{Velocity distribution along $z$ at $t = 0$ (yellow solid bar) and 
at $t= 5\times 10^4 \, \omega_\rmpe^{-1}$ (bar with red boundary).}
\label{temp_incr}
\end{figure}
\begin{figure}
\includegraphics[width= \linewidth]{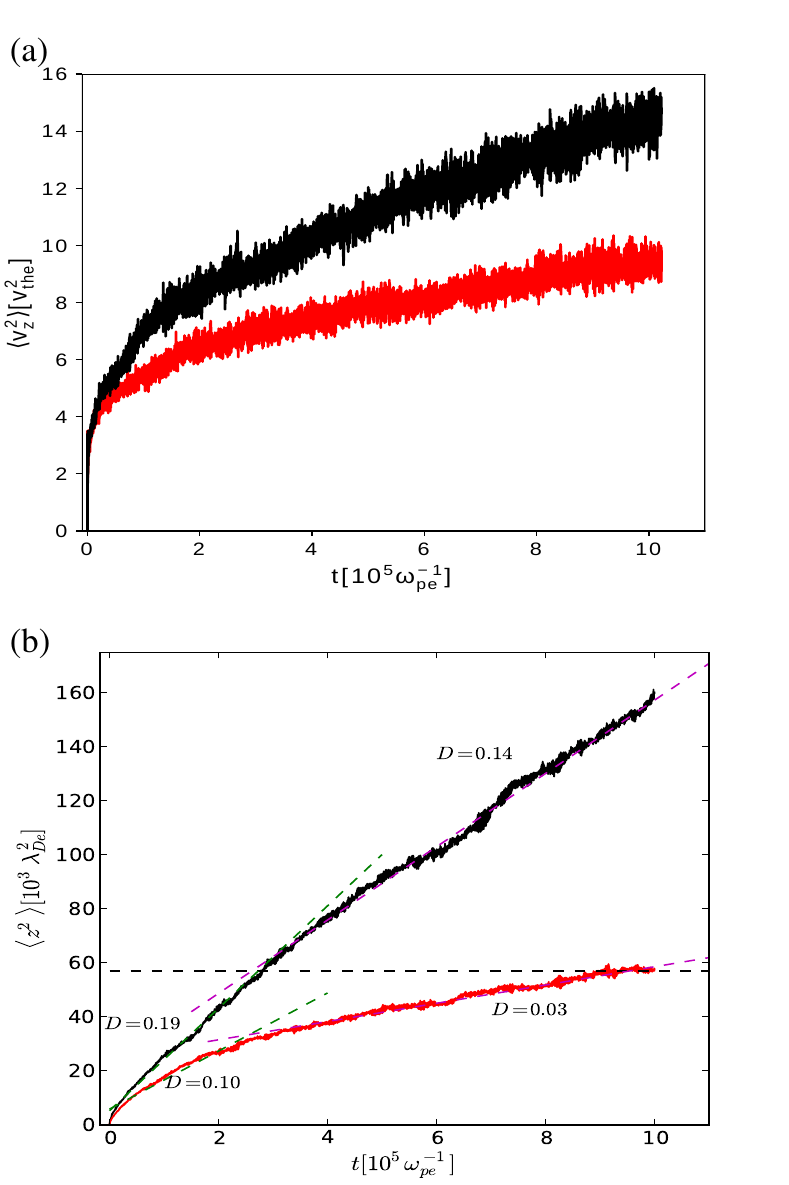}
\caption{Panels (a) and (b): mean square velocity dispersion  
$\langle v_z^2(t)\rangle$  and mean square displacement 
$\langle z^2(t)\rangle $, respectively. The red and black lines correspond 
to no-boundary and reflecting boundary cases, respectively
Panel (b) reveals two diffusion regimes in each curve, 
namely slopes (0.10, 0.03) for no-boundary and (0.18, 0.14) for reflecting 
boundary.}
\label{Trans}
\end{figure}
To analyze the transport, we consider 1056 particles with random initial 
positions in the rectangle 
$0 \leq x_0 \leq 2\pi/ k_{1x}$, $0\leq y_0 \leq 4\pi/ k_{1y}$, $z_0 = 0$ 
and with velocities drawn from a 3D Gaussian 
distribution with unit standard-deviation (viz.\ the thermal velocity) along all three directions. 
Then we evolve their dynamics in the presence of all three waves with equal 
amplitude $\phi_{n0} = {\phi_{0, \mathrm{rms}}}$. For single wave 
interaction, the Hamiltonian of the dynamics can be written in a time 
independent form and therefore, though the dynamics remains chaotic, 
there is no net gain/loss of energy over long time evolution. 
Hence, due to the chaotic dynamics, in presence of the single wave, 
we get a very small amount of cross-field transport along the $z$ direction, 
but the diffusion coefficient is very small. 

But in presence of two or more waves, the Hamiltonian is no longer time 
independent, all the trajectories become chaotic and, due to the wave-particle 
interaction, they gain energy from the waves. 
The particles net perpendicular velocity components $v_y, v_z$ increase. 
After a sufficiently long time-evolution, they form a Gaussian-like velocity 
distribution profile with higher temperature along $y$- and $z$-directions. 
Since $E_x \ll E_{y,z}$, the increase of the velocity component along the 
magnetic field is negligible compared to the other two directions. 
Therefore, the temperature along the magnetic field remains nearly unchanged. 
Fig.~\ref{temp_incr} presents the initial ($t=0$) (solid yellow bars) 
and final $(t= 5\times 10^4 \, \omega_\rmc^{-1})$ (bars with red border) 
velocity distribution of $v_z$, which presents a 
significant increase of temperature along perpendicular direction $T_\perp$ 
compared to the parallel direction, $T_{\perp} / T_{\parallel} \sim 4$.

In the thruster chamber, there is an insulating boundary along $x$-direction. 
The width of the annular space in the thruster is $240 \, \lambda_\rmDe$. 
Therefore the particles are reflected when they reach the boundary. 
If there were no reflection, particles would proceed under 
the same dynamics (red line in Fig.~\ref{Trans}(a)-(b)). 
To account for reflection (black line), we consider the Debye sheath electron 
potential energy near the wall \cite{daren:yu} to be 
${\phi_{\mathrm{sh}} = 20 \,{\mathrm{eV}} = 0.8 \, m_\rme v_{\mathrm{the}}^2}$. 
Electrons reaching the wall with $v_x < \sqrt{0.8}$ are 
specularly reflected, and electrons with $v_x > \sqrt{0.8}$ are isotropically 
reflected from the wall while conserving their total energy. 

Fig.~\ref{Trans}(a)-(b) present $\langle v_z^2(t)\rangle $ and 
$\langle z^2(t)\rangle $ for reflecting boundary (black) and without boundary 
(red), where $\langle \cdot \rangle$ denotes the average over number of particles
for the \emph{deviation from the ballistic motion}. 
Thus, $\langle z^2 \rangle:= \langle (z(t) - v_{z0} t)^2 \rangle$,
$\langle v_y \rangle:= \langle (v_y(t) - v_{y0}) \rangle$,
$\langle v_z \rangle:= \langle (v_z(t) - v_{z0}) \rangle$
 and $\langle v_z^2 \rangle:= \langle (v_z(t) - v_{z0})^2 \rangle$.
The duration of strong interaction with the waves and hence the 
gain of energy from the waves decrease for larger particle velocity.
Therefore, the rate of energy gain in Fig.~\ref{Trans}(a) decreases with time 
for both cases. 
\begin{figure}[t]
\includegraphics[width=\linewidth]{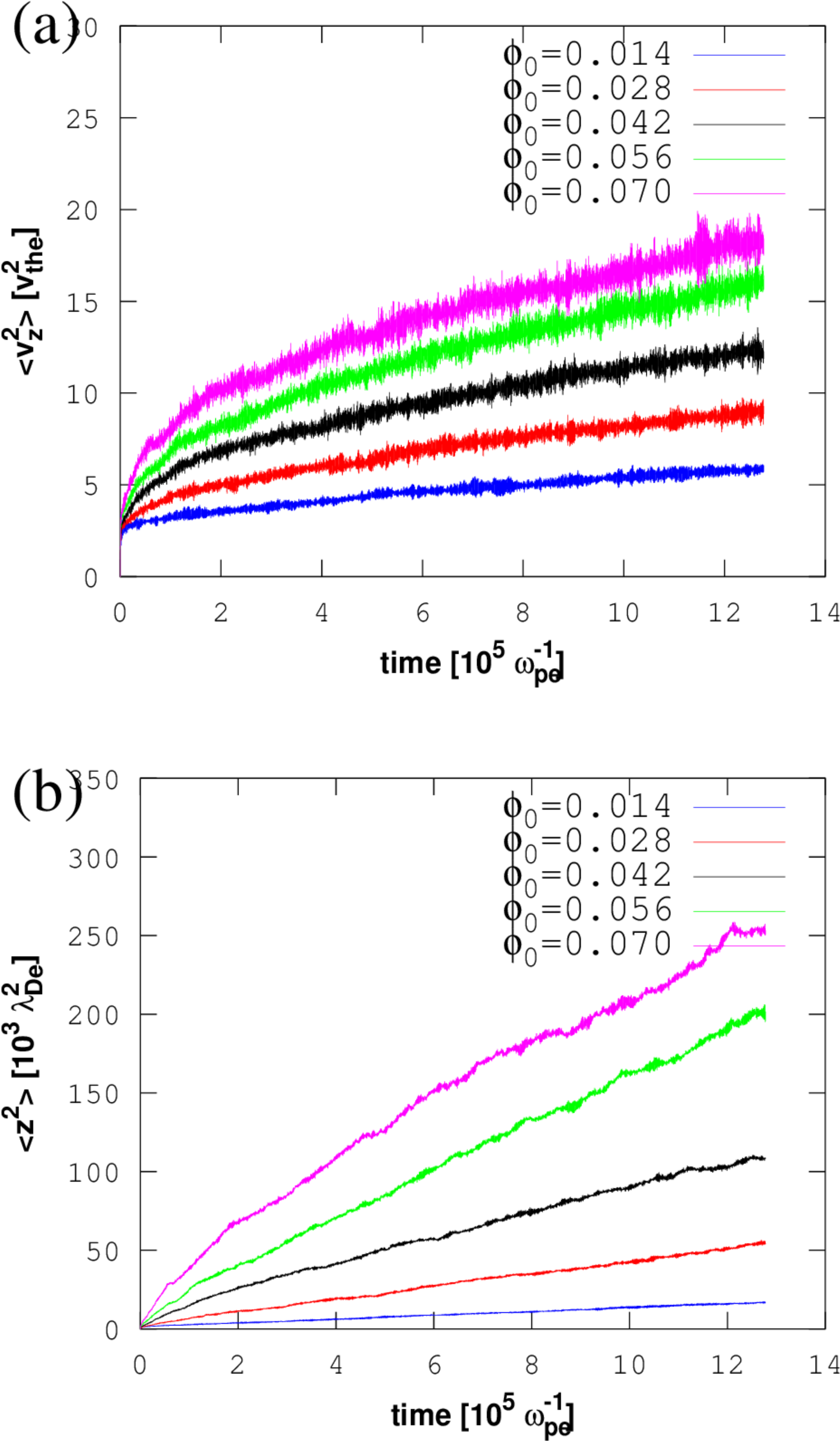}
\caption{Panel (a): $\langle v_z^2 \rangle$ evolution for different amplitude 
of the waves. Panel (b): $\langle z^2 \rangle$ evolution for different wave 
amplitudes.}
\label{amplitude_var}
\end{figure}
\begin{figure}[t]
\includegraphics[width=\linewidth]{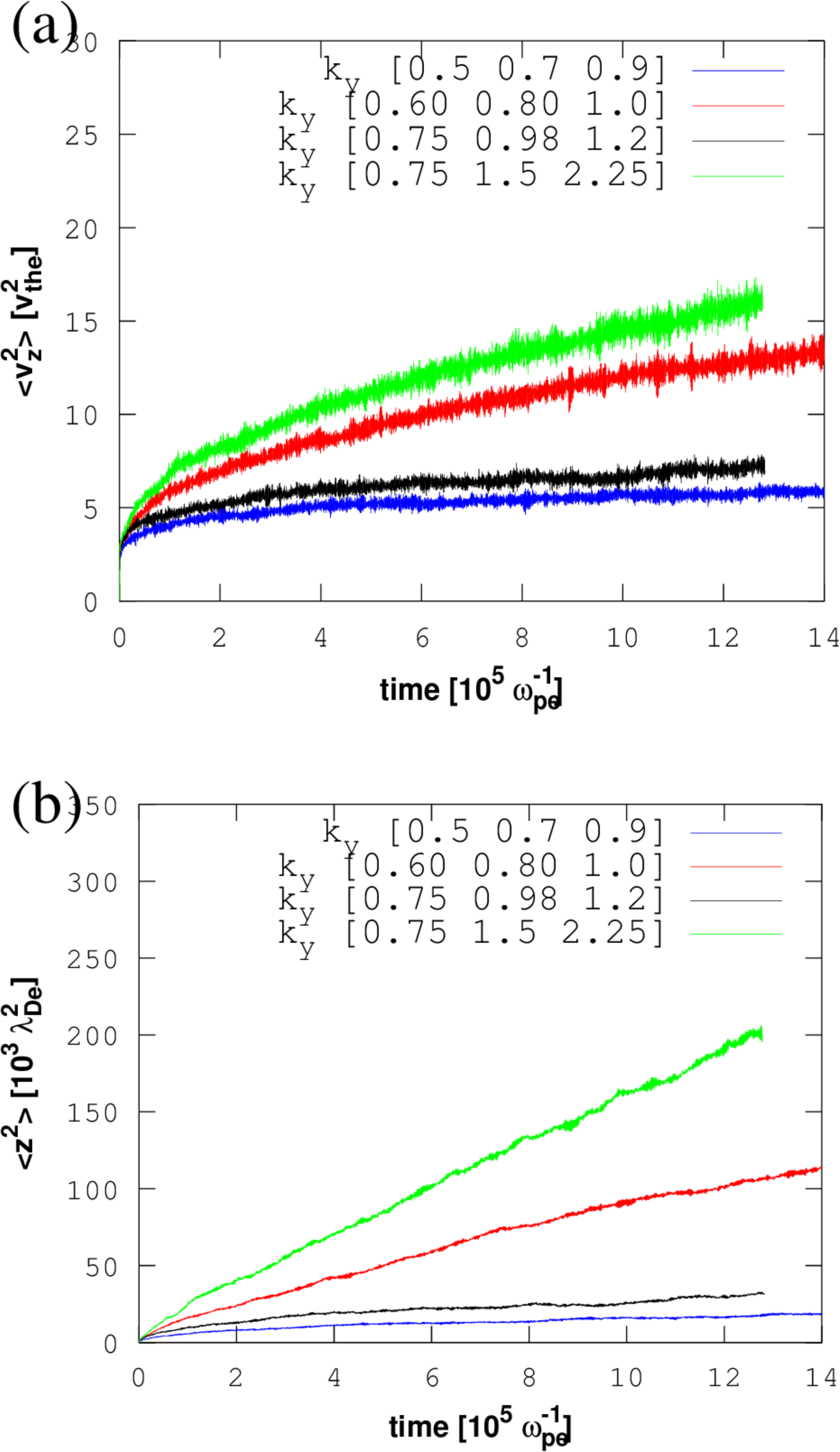}
\caption{ Panel (a): $\langle v_z^2 \rangle$ evolution for different $k_y$ 
values of the waves. Panel (b): $\langle z^2 \rangle$ evolution for different 
$k_y$ values waves.}
\label{k_variation}
\end{figure}

In isotropic reflection, the velocity components of the particle are 
redistributed randomly in three directions, a particle with small 
$v_y$ and $v_x$ gains more energy from the electrostatic wave compared to 
that having higher $v_y$ and $v_x$. Therefore, in presence of reflecting 
boundary, particles gain more energy than in absence of reflection. The 
dashed black line marks the location of thruster outlet along the 
$z$-direction. Since with reflection they gain more energy, their mean square 
displacement along $z$-direction crosses the thruster outlet, and they exit 
from the thruster chamber more quickly than in the case without boundary. 
For both cases, we found two different regimes of transport.
Although the particle motion is not brownian, one may define 
an effective diffusion coefficient 
$D = \rmd \langle z ^2 \rangle / \rmd t$
in the direction of the static electric field $\bf E_0$,
as the average of the slope of $\langle z^2 \rangle$ as a function of time. 
While the derivative $\rmd \langle z ^2 \rangle / \rmd t$ 
is fluctuating strongly, the trend is quite 
stable over time spans on the order of $10^5 \, \omega_{\rmpe}^{-1}$. 
Its observed values are $D = (0.1, 0.03)$ for no-reflection 
and $D = (0.18, 0.14)$ for reflecting boundary. The change in slope around 
$t = 2 \times 10^5 \, \omega_\rmpe^{-1}$ is 
related to the different structure formation of the stochastic web,
controlling the velocity transport \cite{zaslavsky,leoncini}.

The cross-field transport and the energy gain by the particles depend on 
the duration of wave-particle interaction determined by the ratio 
$\omega_\rmb / \omega_\rmc$. Since $\omega_\rmb\propto \sqrt{\phi_0}$, 
the cross-field transport and the energy gain from the wave will be higher 
for higher amplitude of the background waves. Figs~\ref{amplitude_var}(a) and 
(b) present the time evolution of $\langle v_z^2 \rangle$ and 
$\langle z^2 \rangle$ for five different amplitudes of the waves 
$\phi_0 = 0.014,$ $0.028$, $0.042$, $0.056$ and $0.070$. 

Moreover, the strength of the electric field and the bounce frequency 
$\omega_\rmb$ are proportional to the wave number $k_y$, 
therefore $\langle v_z^2 \rangle$ and $\langle z^2 \rangle$ increase for 
larger $k_y$ of the three waves. Fig.~\ref{k_variation}(a) and (b) present 
$\langle v_z^2 \rangle$ and $\langle z^2 \rangle$ for four different sets of 
$k_y$ values, $(k_{y1}, k_{y2}, k_{y3}) = (0.5, 0.7, 0.9); (0.6, 0.8, 1); 
(0.75, 0.98, 1.2); (0.75, 1.5, 2.25)$. 
For the 2nd and 4th cases, the $k_y$ values are the harmonics of $k_y=0.2$ 
and $0.75$, respectively. The large values of $\langle v_z^2 \rangle$ and 
$\langle z^2 \rangle$ for these two cases are due to the formation of 
stochastic webs, which help in long range transport \cite{vasilev}. For the 
other two cases 
where the $k_y$ values are not harmonics, there is no stochastic web 
formation, so that $\langle v_z^2 \rangle$ and $\langle z^2 \rangle$ are 
small for those two cases. 
\begin{figure}
\includegraphics[width=\linewidth]{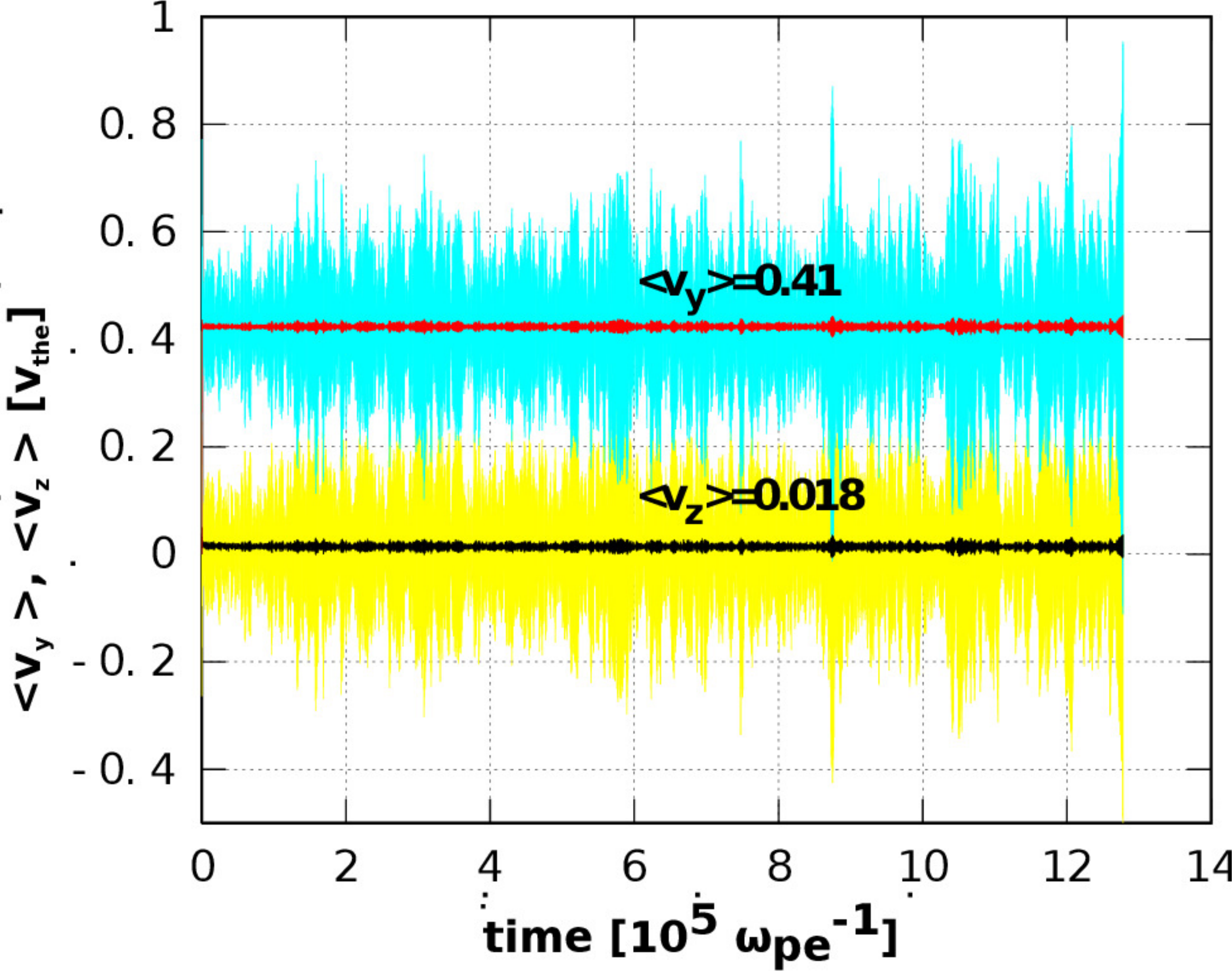}
\caption{Time evolution of $\langle v_y \rangle$ (cyan) and 
$\langle v_z \rangle$ (yellow) of 1056 particles. The solid red line and the 
black line present the time average of the 
$\langle v_y \rangle$ and $\langle v_z \rangle$.}
\label{current_ratio}
\end{figure}

In Hall thrusters, it is experimentally observed \cite{Janes} that 
the ratio of azimuthal to axial current-density $J_y / J_z \sim 10$. In our 
numerical study, we use 1056 particles with random initial
positions in the rectangle 
$0 \leq x_0 \leq 2\pi/ k_{1x}$, $0\leq y_0 \leq 4\pi/ k_{1y}$, $z_0 = 0$ 
and with velocities drawn from a 3D Gaussian
distribution with unit standard-deviation along all three directions. 
We observe that the velocity distributions along 
$z$- and $y$-directions are nearly identical, therefore the number densities 
along these two directions are equal. We can compare the mean velocity
ratio with the mean current density ratio along these two directions. 
Fig.~\ref{current_ratio} shows that $\langle v_y \rangle= 0.41$ and 
$\langle v_z \rangle = 0.018$, so that 
$\langle v_y \rangle / \langle v_z \rangle \sim 20$, 
which is of the same order as the experimental observation.  
%
\section{Conclusions}
\label{sec:conclusion}
%
In this paper, we carried out model calculations to provide a dynamical 
basis for the high value of the experimentally observed anomalous cross-field 
transport in thruster configurations. The underlying mechanism is associated 
with the chaotic dynamics of electrons due to their interaction with a 
spectrum of unstable electrostatic waves. The electrostatic waves are 
generated due to the $\bfE \times \bfB$ electron drift instability. 
In the presence of a magnetic field $B_0$, an axial constant electric 
field $E_0$ and the electrostatic waves, the drifted cyclotron motion becomes 
chaotic due to the strong wave-particle interaction. In presence of more than 
one wave, the electrons gain energy over long time evolution 
and their temperature is increased along the perpendicular direction. 
This chaotic dynamics helps in the transport of electrons along the thruster 
axial direction. 

A significant amount of axial electron transport is observed in presence of 
more than one wave, and the electrons exit from the thruster chamber. 
The reflection at boundary enhances the transport coefficient. 
The duration of wave-particle interaction depends on the ratio 
$\omega_\rmb / \omega_\rmc$ of bounce frequency to cyclotron 
frequency. With increase of amplitude and $k_y$ values of the background waves, 
the value of the bounce frequency increases, which enhances the energy 
exchange rate and the anomalous diffusion coefficient. 
The existence of harmonics in $k_y$ helps to generate different stochastic 
webs, which increases the diffusion coefficient. The average velocity ratio 
along azimuthal to axial direction $\langle v_y \rangle / \langle v_z \rangle$ 
in our numerical model is in good agreement with experimental observations.  
%
\section*{Acknowledgements}
%
We acknowledge the financial support from CEFIPRA/IFCPRA through project 5204-3.
This work was granted access to the HPC resources of Aix-Marseille 
Universit{\'e} \cite{mesocentre} financed by the project Equip{$@$}Meso 
(ANR-10-EQPX-29-01) of the program Investissements d'Avenir supervised by the 
Agence Nationale de la Recherche. We are grateful to Professors Xavier 
Leoncini, Dominique Escande and Abhijit Sen for many fruitful discussions and 
their comments. 

\end{document}